# THE EFFECT OF A MAGNETIC FIELD ON THE MOTION OF ELECTRONS FOR THE FIELD EMISSION PROCESS DESCRIPTION

**S.O. Lebedynskyi, V.I. Miroshnichenko, R.I. Kholodov, V.A. Baturin**
*Institute of Applied Physics of National Academy of Sciences of Ukraine,
58, Petropavlivska St., 40000 Sumy, Ukraine
E-mail: lebedynskyi.s@gmail.com*

The Schrödinger equation is solved for the wave function of an electron moving in a superposition of external constant and uniform electric and magnetic fields at an arbitrary angle between the field directions. The changing of the potential barrier under influence of the magnetic field parallel to the metal surface is shown.

## 1. INTRODUCTION

The researchers and designers of accelerating structures for the compact linear accelerator under the CLIC project were faced with high vacuum RF breakdowns accompanying the electromagnetic power input that produces the accelerating field gradients as high as 100 MV/m.

Toward this end, CERN workers have built a DC-spark facility [1] for use in experiments to elucidate how the frequency of a high-vacuum breakdown occurring in the gap between the electrodes is related to various factors, e.g. material of the electrodes in the accelerating gap, electrode surface conditioning procedures, influence of other circumstances (e.g. external magnetic field in the gap between the electrodes, etc.).

The influence of the external magnetic field on the field emission current density was first studied theoretically by F.J. Blatt [2] and later on in experiments [3-9]. The motivation for these works was a goal to determine the ways of increasing the field emission current, so they confined themselves to the examination of a configuration of the collinear electric and magnetic fields.

However, the central issue, viz, the determination of the electron barrier-penetration coefficient at the metal-vacuum interface was dealt with by the author [2], as he himself admitted, under the assumption that this coefficient was independent of the magnetic field in the configuration of the collinear electric and magnetic fields of interest. Moreover, he advanced neither theoretical arguments nor experimental evidence to support his assumption. The present paper is an attempt to make a first step towards the elucidation of this point, namely, to describe the quantum-mechanical motion of an electron in external constant and uniform electric and magnetic fields, with the angle between the field directions being arbitrary.

It is worthy of note that a study of the magnetic field effect on the field emission is important to perform not only with the aim to prevent high vacuum breakdowns occurring in modern colliders, but also to probe a wide area extending from astrophysics observations of the electron field emission from magnetized neutron stars [10-12] to investigations into the field emission in carbon nanotubes [13-14].

## 2. PROBLEM FORMULATION AND SOLUTION

In our treatment, to describe the electron quantum-mechanical motion we choose a Cartesian system depicted in Fig. 1, with the electric field strength vector $\vec{E_0}$ and the magnetic field induction vector $\vec{B_0}$ directed as indicated in the figure.

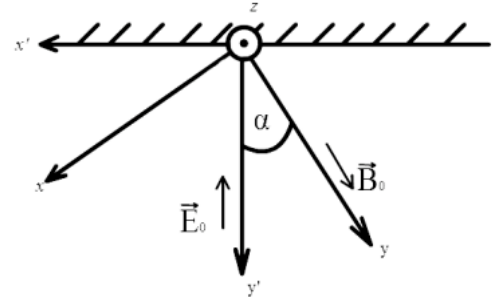

*Fig 1. Configuration of the electric and magnetic fields*

As is seen, the components of the electric field strength $\vec{E_0}$, of the magnetic induction $\vec{B_0}$, the form of the electric potential $\varphi(\vec{r})$, the electron potential energy and the expression for the vector potential can be written as

$$\vec{E_0}(-E_0\sin\alpha, -E_0\cos\alpha, 0),$$
$$\vec{B_0}(0, B_0, 0),$$
$$\varphi(x,y) = E_0(x\sin\alpha + y\cos\alpha), \quad (1)$$
$$U(x,y) = -e\varphi(x,y) = -eE_0(x\sin\alpha + y\cos\alpha),$$
$$\vec{A} = (0, 0, -B_0 x).$$

The input equation for the description of the electron motion in the fields mentioned above is the Schrödinger equation for the electron wave function $\psi(\vec{r},t)$:

$$i\hbar \frac{\partial \psi(\vec{r},t)}{\partial t} = \hat{H}\psi(\vec{r},t), \quad (2)$$

where the Hamilton operator is

$$\hat{H} = \frac{1}{2m}\left[-\hbar^2 \frac{\partial^2}{\partial x^2} - \hbar^2 \frac{\partial^2}{\partial y^2} + \left(-i\hbar \frac{\partial}{\partial z} - eB_0 x\right)^2\right] - eE_0(x\sin\alpha + y\cos\alpha), \quad (3)$$

and $m$ is the electron mass, $-e$ is the electron charge.

As follows from the explicit form of the Hamiltonian operator $\hat{H}$ (3), it does not explicitly depend on time and does not include the $z$ – coordinate in the explicit form, so it is commutative with the operator of the $z$-th component of the momentum we can get the equation for the wave function component $\tilde{\psi}(x,y)$ that describes the electron motion in the $(x,y)$ plane

$$\left\{\frac{1}{2m}\left[-\hbar^2 \frac{\partial^2}{\partial x^2} - \hbar^2 \frac{\partial^2}{\partial y^2} + (p_z - eB_0 x)^2\right] - eE_0(x\sin\alpha + y\cos\alpha)\right\} \times \tilde{\psi}(x,y) = \varepsilon\tilde{\psi}(x,y). \quad (4)$$

The differential operator in Eq. (4) is an algebraic sum of two operators, each depending on only one variable either x or y

$$\hat{H}(x,y) = \hat{H}_x(x) + \hat{H}_y(y), \quad (5)$$
$$\hat{H}_x(x) = \frac{1}{2m}\left[-\hbar^2 \frac{\partial^2}{\partial x^2} + (p_z - eB_0 x)^2\right] - eE_0 x \sin\alpha, \quad (6)$$
$$\hat{H}_y(y) = -\frac{\hbar^2}{2m}\frac{\partial^2}{\partial y^2} - eE_0 y \cos\alpha, \quad (7)$$
$$\{\hat{H}_x(x) + \hat{H}_y(y)\}\tilde{\psi} = \varepsilon\tilde{\psi}(x,y), \quad (8)$$
$$\varepsilon_x + \varepsilon_y = \varepsilon, \quad (9)$$

with the constants $\varepsilon_x$ and $\varepsilon_y$ determine the possible spectrum of the electron energy related to its movement along either $x-$ or $y-$ axis.

Considering the additive nature of the Hamiltonian operator with respect to the dependence on the $x, y$ coordinates, we search for the electron wave function, $\tilde{\psi}(x,y)$ in the form of a product of two functions, each being dependent on only one variable

$$\tilde{\psi}(x,y) = X(x)Y(y). \quad (10)$$

To define the X(x) function the quantum oscillator differential equation is derived

$$\frac{d^2 X(\xi)}{d\xi^2} + (\tilde{\varepsilon}_x - \xi^2)X(\xi) = 0, \quad (11)$$

where we introduce a dimensionless coordinate ξ in accordance with the expressions

$$x - x_0 = x', x_0 = \frac{p_z}{m\omega_B} + \frac{eE_0 \sin\alpha}{m\omega_B^2}, a = \left(\frac{\hbar}{m\omega_B}\right)^{1/2},$$
$$x' = a\xi, \xi = \left(\frac{m\omega_B}{\hbar}\right)^{1/2}\left(x - \frac{p_z}{m\omega_B} - \frac{eE_0 \sin\alpha}{m\omega_B^2}\right), \quad (12)$$

$\omega_B = \frac{eB_0}{m}$ is the cyclotron frequency of the electron rotation in Larmor orbit in the classical case and the constant $\tilde{\varepsilon}_x$ is represented by the expression

$$\tilde{\varepsilon}_x = \frac{1}{\hbar\omega_B}\left[2\varepsilon_x + \frac{e^2 E_0^2 \sin^2\alpha}{m\omega_B} + \frac{2eE_0 \sin\alpha}{m\omega_B}p_z\right]. \quad (13)$$

Equation (11) is intended for Hermitian functions [15] and has a finite solution only for a discrete series of $\tilde{\varepsilon}_x$ values:

$$\tilde{\varepsilon}_x = (2n+1), \text{where } n = 0,1,2,\ldots \quad (14)$$

The expressions (13 and 14) determine a possible range of energy $\varepsilon_x$ which is related to the electron motion in the plane normal to the magnetic induction vector and can be represented as

$$\varepsilon_x = \left(n + \frac{1}{2}\right)\hbar\omega_B + \frac{mv_d^2}{2} - eE_0 x_0. \quad (15)$$

$v_d = \frac{E_0}{B_0}\sin\alpha$ is a drift velocity of the electron Larmor orbit center in crossed electric and magnetic fields in the classical description.

The part of the electron wave function that describes the electron motion in the plane normal to the magnetic field, can be represented in the explicit form as

$$\psi(x,z) = \frac{1}{\sqrt{\pi 2^n n!}} exp\left(-\frac{\xi^2}{2}\right) H_n(\xi) exp\left(\frac{ip_z z}{\hbar}\right), \quad (16)$$

where $H_n(\xi)$ is the Hermitian polynomial of n-th order.

Equation (7) that defines the electron motion along the magnetic field can be reduced to

$$\frac{d^2 Y}{d\eta^2} - \eta Y = 0, \quad (17)$$

where the dimensionless η coordinate is linked to the y coordinate by the following relation

$$\eta = \left(\frac{2meE_0 \cos\alpha}{\hbar}\right)^{1/3}\left(y - \frac{E_y}{eE_0 \cos\alpha}\right). \quad (18)$$

The solution of Equation (17) can be reduced to the solution of the equation for the Bessel function of order 1/3 [15]:

$$Y(\eta) = AH_{\frac{1}{3}}^{(1)}(\eta) + BH_{\frac{1}{3}}^{(2)}(\eta), \quad (19)$$

where $H_{\frac{1}{3}}^{(1)}(\eta)$ and $H_{\frac{1}{3}}^{(2)}(\eta)$ are the Hankel functions of the 1st- and 2d-order, respectively[15]. The range of electron energies $\varepsilon_y$ is the characteristic of the electron motion along the magnetic field; it assumes a continuous series of values.

## 3. BARRIER-PENETRATION COEFFICIENT IN METAL IN THE PRESENCE OF COLLINEAR MAGNETIC AND ELECTRIC FIELDS

We turn here to the problem of the dependence or independence of the barrier-penetration coefficient of a metal electron under electron field emission in the case of parallel electric and magnetic fields $(\vec{E_0}||\vec{B_0})$. Choosing, as before, the form of the magnetic field vector potential to be $\vec{A} = (0,0,-eB_0 x)$ we can write the explicit form of the Hamiltonian operator of an electron penetrating the potential barrier in metal due to the external electric field as follows

$$\hat{H} = \frac{1}{2m}\left[-\hbar^2 \frac{\partial^2}{\partial x^2} - \hbar^2 \frac{\partial^2}{\partial y^2} + \left(-i\hbar\frac{\partial}{\partial z} - eB_0 x\right)^2\right] + \left[C - eE_0(x\sin\alpha + y\cos\alpha) - \frac{e^2}{16\pi\varepsilon_0(x\sin\alpha + y\cos\alpha)}\right] \times \sigma(x\sin\alpha + y\cos\alpha), \quad (20)$$

where $\sigma(\alpha) = \begin{cases} 0 & \alpha < 0 \\ 1 & \alpha \geq 0 \end{cases}$.

Equation (18) implies the same coordinate system as that shown in Fig.1 and includes the following symbols: $C$ is the potential barrier height; $-eE_0(x\sin\alpha + y\cos\alpha)$ is the potential energy of an electron present in the external electric field of given configuration; $e^2/16\pi\varepsilon_0(x\sin\alpha + y\cos\alpha)$ is the electron potential energy related to the interaction of the electron with its "positive" mirror image.

Examine the form of the Hamiltonian operator in a special case of the electric and the magnetic fields being collinear, i.e. at the angle α = 0 (as discussed in [2]). In this case the Hamiltonian operator has additive nature, that is, it can be represented as a sum of two operators, each acting on only one variable, either x or y

$$\hat{H}(x,y) = \hat{H}_x(x) + \hat{H}_y(y), \quad (21)$$
$$\hat{H}_x(x) = \frac{\hbar^2}{2m}\left[-\frac{\partial^2}{\partial x^2} + (p_z - eB_0 x)^2\right], \quad (22)$$
$$\hat{H}_y(y) = -\frac{\hbar^2}{2m}\frac{\partial^2}{\partial y^2} + \left(C - eE_0 y - \frac{e^2}{16\pi\varepsilon_0 y}\right)\sigma(y). \quad (23)$$

Here the Schrödinger equation for the electron wave function $\tilde{\psi}(x,y)$ can be written as

$$\{\hat{H}_x(x) + \hat{H}_y(y)\}\tilde{\psi} = \varepsilon\tilde{\psi}(x,y). \quad (24)$$

Considering the additive nature of the differential operator in (21-23), the expression for the component of the $\tilde{\psi}(x,y)$ wave function is sought for in the multiplicative form

$$\tilde{\psi}(x,y) = X(x)Y(y). \quad (25)$$

Substitution of (24) into (25) makes it necessary to solve two independent differential equations

$$\hat{H}_x(x)X(x) = \varepsilon_x X(x), \ x \in (-\infty,+\infty), \quad (26)$$
$$\hat{H}_y(y)Y(y) = \varepsilon_y Y(y), \ y \in (-\infty,+\infty). \quad (27)$$

where the quantities $\varepsilon_x$ and $\varepsilon_y$ determine possible electron energy ranges along the magnetic field and in the perpendicular thereto plane. The relations (26) and (27) reveal a possibility of an independent description of the electron motion along the $x-$ and $y-$ axis as the electron penetrates the potential barrier at the metal-vacuum interface in the case of the collinear electric and magnetic fields. Proceeding from the above considerations, we can conclude that the barrier-penetration coefficient is independent of the magnetic field for the case of parallel electric and magnetic fields as was rightly supposed but not proved in [2].

## 4. THE CHANGING OF THE POTENTIAL BARRIER UNDER THE MAGNETIC FIELD INFLUENCE

To better understand the influence of the external constant magnetic field on the field emission process let's consider its effect on the form of the potential step at the metal-vacuum surface in the case when the magnetic field is parallel to the surface. The Schrödinger equation for this case takes the following form

$$\left\{\frac{1}{2m}\left[-\hbar^2\frac{d^2}{dx^2}+p_y^2+(p_z-eB_0x)^2\right]+\left[-eE_0x-\frac{e^2}{16\pi\varepsilon_0 x}\right]\right\}\psi(x)=\varepsilon\psi(x), \quad (28)$$

where the Hamilton operator is

$$\widehat{H}=\frac{1}{2m}\left[-\hbar^2\frac{\partial^2}{\partial x^2}--\hbar^2\frac{\partial^2}{\partial y^2}+\left(-i\hbar\frac{\partial}{\partial z}-eB_0x\right)^2\right]-eE_0x-\frac{e^2}{16\pi\varepsilon_0 x}.$$

As follows from the explicit form of the Hamiltonian operator $\widehat{H}$, it does not explicitly depend on time and does not include the $y-$ and $z-$ coordinates in the explicit form, so it is commutative with the operator of the *y*-th and *z*-th components of the momentum we can get the equation for the wave function component $\psi(x)$ that describes the electron motion

$$\frac{d^2\psi}{dx^2}-\frac{2m}{\hbar^2}\left[\frac{\omega_B^2 x^2 m}{2}-eE_0x-\frac{e^2}{16\pi\varepsilon_0 x}-\varepsilon'\right]\psi=0, \quad (29)$$

where $\varepsilon'=\varepsilon+\frac{p_y^2}{2m}+\frac{p_z^2}{2m}$. And the effective potential energy $V(x)$ of an electron near a metal surface is described as following

$$V(x)=\frac{\omega_B^2 x^2 m}{2}-eE_0x-\frac{e^2}{16\pi\varepsilon_0 x}. \quad (30)$$

Fig.2 in different scales shows a comparison of the potential barrier near the metal surface in the absence of a magnetic field and in the case of an external uniform magnetic field parallel to the surface of the metal.

From the figure we can see that near the surface the form of potential step remains intact. But at some distance from the metal surface the potential step becomes infinite. As result we expect that the field emission process in presence of the external magnetic field parallel to the metal surface will occur only for a limited interelectrode distance.

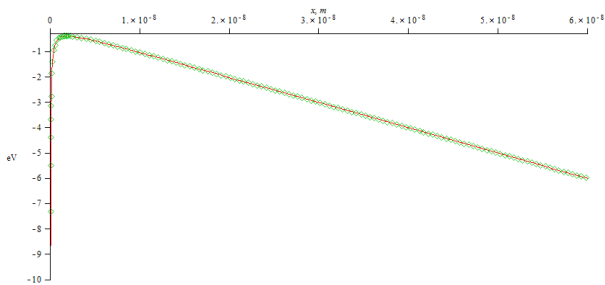

a)

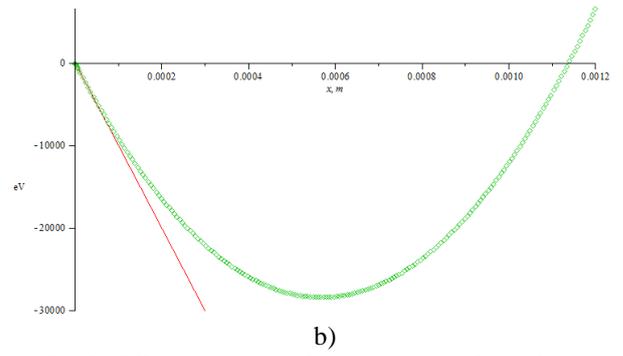

b)

*Fig.2. Effective potential energy V(x) of an electron near a metal surface, as given by eq. (30) in different scales: a) from 0 to 6·10⁻⁸ m, b) from 0 to 1.2·10⁻³ m. The line is the case of E=100 MV/m, B=0.*
*By dots shows the case E=100 MV/m, B=1 T.*

## 5. THE POSSIBLE EXPERIMENTS

The experimental studying of the effect of the magnetic field on the field emission current is planned in Institute of Applied Physics, National Academy of Sciences of Ukraine. The fig. 3 shows the experimental setup, which is built to study the high vacuum high gradient breakdowns, but it can operate for researching the field emission current .

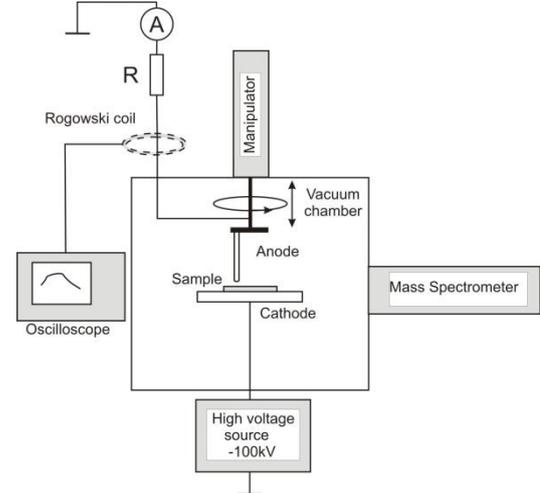

*Fig.3. Schematic drawing of the experimental setup.*

The composition of the experimental setup consists of: high-vacuum chamber with the sample-fixing mechanism that allows to control motion of the samples during the experiment, monopole mass spectrometer for control the composition of the atmosphere in a vacuum, system of registration the current before breakdown and directly the breakdown, the system of heating vacuum chamber and computer control system installation.

This setup allows to set the gap from 10 microns to 1 mm and apply voltage up to 50 kV. The setup has all necessary equipment, that allows to measure the field emission current down to 0.1 nA. These parameters allows to investigate current in wide regions of gradients and gaps.

Theoretically field emission current is well described by the Fowler-Nordheim equation which includes image forces gives the following expression for the current density:

$$j = \frac{e^3 E^2}{8\pi h \varphi} \exp\left(\frac{4\sqrt{2m}\, \varphi^{\frac{3}{2}}}{3\hbar e E}\, v\left[\frac{\sqrt{e^3 E}}{\varphi}\right]\right), \quad (31)$$

where $\varphi$ is the work function of electrons, $v(y)$ is Nordheim function that has been evaluated for representative values of *y*.

The explicit form of the expression for the field emission current-density contains the electric field strength and we assume that it is possible to find evaluation of the influence of the magnetic field using Lorentz covariance. For case the same electromagnetic invariants in the presence of the magnetic field the electric field strength will change as following

$$E^* = E\sqrt{1 - \frac{B^2 c^2}{E^2}}. \quad (32)$$

As can be seen from the equation the influence of the electric field should be reduced in presence of the magnetic field.

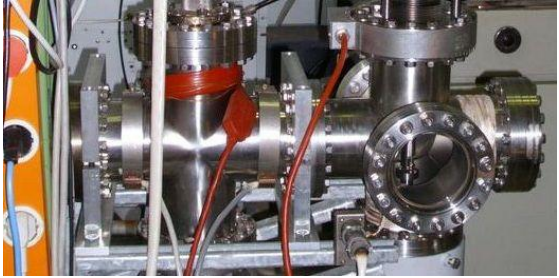

*Fig.4. The working chamber of the experimental setup.*

The working chamber of the experimental setup (Fig. 4) allows puts inside the magnet for studying the magnetic field influence on the field emission current. It is possible to conduct a study the influence of the magnetic field magnitude of 1.5 Tesla. According to the preliminary estimates (eq. 32) for the field emission current order of hundreds nA the influence of the magnetic field will be about 20% of current without magnetic field. Studying the possibility of locking the field emission current by external magnetic field also exists in this experimental setup.

## 6. SUMMARY

The authors propose a solution for the problem of electron quantum-mechanical motion in external constant and uniform electric and magnetic fields crossing at an arbitrary angle. The electron wave function has been derived in the explicit form for an electron moving in thus superimposed fields.

It is shown that in the case of collinear electric and magnetic fields the coefficient of the potential barrier penetration by the electron does not depend on the magnetic field. This fact supports the supposition made by F.J. Blatt [2] which, to our knowledge, has so far received neither theoretical no experimental confirmation.

As is apparent from the form of differential equation the electron barrier penetration coefficient depends in the general case on the magnetic field.

The form of the potential step at the metal-vacuum surface in the case when the magnetic field is parallel to the surface is shown. Hence the field emission current can be controlled by the external magnetic field.

The preliminary estimates for the field emission current under the influence of the magnetic field were done.

The estimation of the barrier penetration coefficient and field emission current in the presence of the magnetic field will be a subject of further investigations.

## ACKNOWLEDGMENTS

Publication is based on the research provided by the grant support of the State Fund For Fundamental Research (project N Ф58/174-2014) as well as by the National Academy of Sciences of Ukraine (NASU) under the program of cooperation between NASU, CERN and JINR Prospective Research into High-Energy and Nuclear Physics under Contract No ЦО-5-1/2014).

**ВЛИЯНИЕ МАГНИТНОГО ПОЛЯ НА ДВИЖЕНИЕ ЭЛЕКТРОНОВ ДЛЯ ОПИСАНИЯ ПРОЦЕССА АВТОЭЛЕКТРОННОЙ ЭМИССИИ**
**С.А. Лебединский, В.И. Мирошниченко, Р.И. Холодов, В.А. Батурин**

Решается уравнение Шредингера для волновой функции электрона, движущегося в суперпозиции внешних постоянных и однородных электрическом и магнитном полях под произвольным углом между их направлениями. Показано изменение потенциального барьера под влиянием магнитного поля, параллельного поверхности.

**ВПЛИВ МАГНІТНОГО ПОЛЯ НА РУХ ЕЛЕКТРОНІВ ДЛЯ ОПИСУ ПРОЦЕСУ ПОЛЬОВОЇ ЕМІСІЇ**
**С.О. Лебединський, В.І. Мирошніченко, Р.І. Холодов, В.А. Батурін**

Розв'язується рівняння Шрьодінгера для хвильової функції електрона, що рухається в суперпозиції зовнішніх постійних і однорідних електричному і магнітному полях під довільним кутом між їх напрямами. Показано зміну потенційного бар'єру під впливом магнітного поля, паралельного поверхні.